\documentstyle[12pt,leqno]{article}
\begin{document}
\title{A NEW CONTRIBUTION TO THE FLAVOUR-CHANGING LEPTON-PHOTON VERTEX}
\author{D. PALLE \\Department of Theoretical Physics,Rugjer Bo\v skovi\'c 
Institute \\P.O.Box 1016,Zagreb,CROATIA}
\date{ }
\maketitle
{\bf Summary.} - 
We show that the correct perturbation theory for mixed fermion 
states leads to nonvanishing contributions of the dimension-four
vertex operators for flavour-changing transitions. Their contributions
to the amplitude are
of the same order of magnitude as the dimension-five vertex operators.
Considerations are valid irrespective of the
electroweak model.\hspace*{12 mm}12.15.Ff Quark and 
lepton masses and mixing \\ \hspace*{12 mm} 13.40 Hq Electromagnetic
 decays
\newpage

It seems that current measurements in astro- and particle physics (solar 
and atmospheric neutrinos, COBE data, LSND data, observed ionization of the
 Universe, etc.) strongly suggest that neutrinos
should be massive particles.  
 In this paper we show that the correct treatment 
of the flavour mixing of massive leptons in perturbation theory results in new
terms in the decay amplitude due to the dimension-four operators.

Contributions to flavour-changing radiative decays of leptons
in the perturbative
calculations of the majority of electroweak models appear through
quantum loops owing to the existence of flavour-changing charged weak 
currents.
We can write the general form of the $f_{1}\rightarrow f_{2}+\gamma$ amplitude
(vertex) with the following Lorentz structures \cite{Moha}:

\begin{eqnarray}
{\cal M}_{\mu}(f_{1}(p_{1})\rightarrow f_{2}(p_{2})+\gamma(q))=-\imath
\bar{u}(p_{2})\Gamma_{\mu ;f_{1}f_{2}}u(p_{1}) \nonumber \\
=-\imath\bar{u}(p_{2})[\gamma_{\mu}(F_{1}^{L}(q^{2})P_{L}+F_{1}^{R}(q^{2})P_{R})
+\imath\sigma_{\mu\nu}q^{\nu}(F_{2}^{L}(q^{2})P_{L}+F_{2}^{R}(q^{2})P_{R})
\nonumber \\
+q_{\mu}(F_{3}^{L}(q^{2})P_{L}+F_{3}^{R}(q^{2})P_{R})]u(p_{1}) , \\
where:\ P_{L,R}\equiv\frac{1}{2}(1\mp\gamma_{5})\ . \nonumber
\end{eqnarray}

Previous calculations \cite{Moha}
of the decay amplitude were focused on the $F_{2}^{L,R}$
form factors  because the $F_{3}^{L,R}$ form factors gave zero 
contribution to $f_{1}\rightarrow f_{2}+\gamma $. However, the 
$F_{1}^{L,R}$ form factors were claimed to give contributions that should vanish
because of the conservation of the electromagnetic current. We show that the
latter claim is incorrect and that the $F_{1}^{L,R}$ form factors give
contributions to the amplitude comparable with that of the $F_{2}^{L,R}$
 \cite{Moha}.

The most natural choice for the renormalization scheme in electroweak
theory is the on-shell renormalization scheme \cite{Aoki}. In this scheme,
we repeat the most important ingredients concerning mixed fermion 
states. The on-shell renormalization conditions for mixed fermions
(propagators) are

\begin{eqnarray}
      S_{ij}^{ren}[pole]=\frac{\delta_{ij}}{m_{i}-\not p}, 
\ \ \ \ \ \ \ \ \ \ \ \ \ \ \ \ \\
\ \ \ where:\ S_{ij}\equiv Fourier\ Transform\  \imath\langle 0|T\psi_{i}
(x)\bar{\psi}_{j}(y)|0\rangle .\nonumber
\end{eqnarray}

These renormalization 
 conditions ensure a correct input for fermion masses and
a correct form of renormalized propagators.
These conditions can be written in a more transparent
form by introducing on-shell spinors \cite{Aoki}:

\begin{eqnarray}
     K_{ij}^{ren}u(m_{j})=0 ,\ \ \ \ \ \ \ \ \ \  \nonumber \\
     \bar{u}(m_{i})K_{ij}^{ren}=0,\ \ \ \ \ \ \ \ \ \  \nonumber  \\
    \{\frac{1}{m_{i}-\not p}K_{ii}^{ren}\}u(m_{i})=u(m_{i}),  \\
   \bar{u}(m_{i})\{K_{ii}^{ren}\frac{1}{m_{i}-\not p}\}=\bar{u}(m_{i}), 
  \nonumber \\
  definition:\ K_{ij}^{ren}(p)\equiv\{S^{ren}(p)\}_{ij}^{-1},\ 
i,j=flavour\ indices. \nonumber
\end{eqnarray}

For Majorana fields, one obtains the same formulae for on-shell renormalization
conditions, but with a different number of conditions in comparison with
Dirac fermions \cite{Aoki}.

Furthermore, any correctly quantized electroweak theory preserves
the BRST symmetry \cite{Aoki}. Thus, the generalized
Ward-Takahashi identity for the flavour-changing lepton-photon vertex
can be written as

\begin{eqnarray}
q^{\mu}\Gamma_{\mu;ij}^{ren}(p+q,p|q)=-e\Sigma^{ren}_{ij}(p)+
e\Sigma^{ren}_{ij}(p+q) . 
\end{eqnarray}

From the general Lorentz structure of the lepton-photon amplitude one can see
that only the $F_{1}^{L,R}$ form factors are related to flavour       
off-diagonal self-energies through Ward-Takahashi identities\cite{Aoki,Boehm}.
These identities are valid for renormalized Green functions, and the 
on-shell renormalization conditions (3) should be used to uniquely fix finite
terms of self-energies:

\begin{eqnarray}
\Sigma^{ren}_{ij}(p)u(p,m_{j})=0 , \nonumber \\
\bar{u}(p,m_{i})\Sigma^{ren}_{ij}(p)=0 .
\end{eqnarray}

It is now evident that the electromagnetic current remains conserved because
of the on-shell conditions:
 
\begin{eqnarray}
q^{\mu}\bar{u}(p,m_{i})\Gamma^{ren}_{\mu;ij}(p+q,p|q)u(p+q,m_{j})=0 . 
\nonumber
\end{eqnarray}

In addition, we can evaluate the $F_{1}^{L,R}(q^{2})$ form factors from the
on-shell conditions at $q^2=0$. 
Let us write the most general form of
the renormalized self-energy (flavour indices suppressed):

\begin{eqnarray}
\Sigma^{ren}(p)\equiv(\sigma_{1}(p^{2})+\delta Z_{1})\not p P_{L}+
(\sigma_{2}(p^{2})+\delta Z_{2})\not p P_{R} \nonumber \\
+(\sigma_{3}(p^{2})+\delta Z_{3}) P_{L}+
(\sigma_{4}(p^{2})+\delta Z_{4}) P_{R}.
\end{eqnarray}
    
Inserting the above into the Ward-Takahashi identity and setting
$p_{\mu}=0\ and\ q^{2}=0$, we obtain the following expressions
for the form factors:

\begin{eqnarray}
F_{1}^{L}(0)_{f_{1}\neq f_{2}}=e(\sigma_{1}(0)+\delta Z_{1}) , \nonumber \\
F_{1}^{R}(0)_{f_{1}\neq f_{2}}=e(\sigma_{2}(0)+\delta Z_{2}) .
\end{eqnarray}

The four renormalization constants are defined by the on-shell 
renormalization conditions (5) ($f_{i}\ fermion\ has\ a\ mass\ m_{i}$):
      
\begin{eqnarray}
\delta Z_{1}=\frac{1}{m_{1}^{2}-m_{2}^{2}}[-m_{1}^{2}\sigma_{1}(m_{1}^{2})
+m_{2}^{2}\sigma_{1}(m_{2}^{2})+m_{1}m_{2}(\sigma_{2}(m_{1}^{2})-
\sigma_{2}(m_{2}^{2})) \nonumber \\
-m_{2}(\sigma_{3}(m_{1}^{2})-\sigma_{3}(m_{2}^{2}))+m_{1}(\sigma_{4}
(m_{1}^{2})-\sigma_{4}(m_{2}^{2}))] , \nonumber \\
\delta Z_{2}=\frac{1}{m_{1}^{2}-m_{2}^{2}}[m_{1}m_{2}(\sigma_{1}(m_{1}^{2})
-\sigma_{1}(m_{2}^{2}))-m_{1}^{2}\sigma_{2}(m_{1}^{2})+m_{2}^{2}
\sigma_{2}(m_{2}^{2}) \nonumber \\
+m_{1}(\sigma_{3}(m_{1}^{2})-\sigma_{3}(m_{2}^{2}))-m_{2}(\sigma_{4} 
(m_{1}^{2})-\sigma_{4}(m_{2}^{2}))] ,  \\
\delta Z_{3}=\frac{1}{m_{1}^{2}-m_{2}^{2}}[m_{2}m_{1}^{2}(\sigma_{1}
(m_{1}^{2})-\sigma_{1}(m_{2}^{2}))-m_{1}m_{2}^{2}(\sigma_{2}
(m_{1}^{2})-\sigma_{2}(m_{2}^{2}))  \nonumber \\
+m_{2}^{2}\sigma_{3}(m_{1}^{2})-m_{1}^{2}\sigma_{3}(m_{2}^{2})- 
m_{1}m_{2}(\sigma_{4}(m_{1}^{2})-\sigma_{4}(m_{2}^{2}))] , \nonumber \\
\delta Z_{4}=\frac{1}{m_{1}^{2}-m_{2}^{2}}[-m_{1}m_{2}^{2}(\sigma_{1}
(m_{1}^{2})-\sigma_{1}(m_{2}^{2}))+m_{2}m_{1}^{2}(\sigma_{2}
(m_{1}^{2})-\sigma_{2}(m_{2}^{2})) \nonumber \\
-m_{1}m_{2}(\sigma_{3}(m_{1}^{2})-\sigma_{3}(m_{2}^{2}))+m_{2}^{2}
\sigma_{4}(m_{1}^{2})-m_{1}^{2}\sigma_{4}(m_{2}^{2})] . \nonumber
\end{eqnarray}

For definiteness, let us write the interaction Lagrangian with flavour-changing
lepton charged currents with Dirac neutrinos, 
in a form that is valid irrespective of the symmetry-
breaking mechanism :

\begin{eqnarray*}
{\cal L}_{I}=\frac{g}{\sqrt{2}}\sum_{ij}W^{\mu}\bar \nu_{i}U_{ij}\gamma_{\mu}P_{L}l_{j}
+\frac{g}{\sqrt{2}M_{W}}\sum_{ij}\phi^{+} \bar{\nu}_{i}[m_{l_{j}}U_{ij}P_{R}-
m_{\nu_{i}}U_{ij}P_{L}]l_{j}+h.c. ,  \\
\nu=neutrino;\ l=charged\ lepton;\ \phi^{+}=Nambu-Goldstone\ scalar .
\end{eqnarray*}

Then we have to find renormalized neutrino self-energies with the above
interaction Lagrangian and the renormalization conditions (5).

In the 't Hooft-Feynman gauge, one can easily verify that

\begin{eqnarray}
\sigma_{1;2;3;4}(p^{2})=-\imath\frac{g^{2}}{32\pi^{2}}\sum_{l}U_{il}U^{*}
_{ml}\times \{(2+\frac{m_{l}^{2}}{M_{W}^{2}})B_{1}(p^{2};m_{l}^{2},
M_{W}^{2}); \nonumber \\
\frac{m_{\nu_{i}}m_{\nu_{m}}}{M_{W}^{2}}B_{1}(p^{2};m_{l}^{2},M_{W}^{2});
\frac{m_{l}^{2}}{M_{W}^{2}}m_{\nu_{i}}B_{0}(p^{2};m_{l}^{2},M_{W}^{2});
\frac{m_{l}^{2}}{M_{W}^{2}}m_{\nu_{m}}B_{0}(p^{2};m_{l}^{2},M_{W}^{2})\} , 
\nonumber \\
\{m,i,l\}\ are\ flavours\ of\ \{\nu_{1},\nu_{2},charged\ lepton\ l\}. \nonumber
\end{eqnarray}

The scalar functions $B_{0}(p^2)$ and $B_{1}(p^2)$\cite{Boehm} have to be evaluated
for $p^2 \ll (M_{1}-M_{2})^2$, so it would be useful to make an expansion
in the vicinity of $p^2=0$:

\begin{eqnarray}
\frac{\imath}{16 \pi^{2}}\{1;p_{\mu}\}B_{0;1}(p^{2};M_{1},M_{2})=
\int \frac{d^{4}k}{(2 \pi)^{4}}\frac{\{1;k_{\mu}\}}{(k^{2}-M_{1}^{2}+\imath
 \epsilon)((k+p)^{2}-M_{2}^{2}+\imath \epsilon)}, \nonumber \\
B_{0}(p^{2};M_{1},M_{2})=\theta (M_{1},M_{2})+b_{2}p^{2}+b_{4}p^{4}+
{\cal O}(p^{6}), \nonumber \\
B_{1}(p^{2};M_{1},M_{2})=\eta (M_{1},M_{2})+\frac{1}{2}(b_{4}\frac
{M_{2}^{2}-M_{1}^{2}}{2}-b_{2})p^{2}+{\cal O}(p^{4}), \nonumber \\
b_{2}=\frac{1}{2}\frac{M_{1}^{2}+M_{2}^{2}}{(M_{1}^{2}-M_{2}^{2})^{2}}
+\frac{2M_{1}^{2}M_{2}^{2}}{(M_{1}^{2}-M_{2}^{2})^{3}}ln\frac{M_{2}}{M_{1}},
\nonumber \\
b_{4}=\frac{M_{1}^{4}+10M_{1}^{2}M_{2}^{2}+M_{2}^{4}}{6(M_{1}^{2}-
M_{2}^{2})^{4}}-2\frac{M_{1}^{2}M_{2}^{2}(M_{1}^{2}+M_{2}^{2})}
{(M_{1}^{2}-M_{2}^{2})^{5}}ln\frac{M_{2}}{M_{1}}, \nonumber \\
\theta,\eta\ functions\ contain\ ultraviolet\ infinity. \nonumber
\end{eqnarray}

From the above we can evaluate the leading terms of the $F_{1}^{L,R}(0)_{f_{1}
\neq f_{2}}$ form
factors ($m_{2}\ll m_{1}\ and\ m_{l}\ll M_{W}$)

\begin{eqnarray}
F_{1}^{L}(0)_{i \neq m}\simeq -\imath\frac{eG_{F}}{4\sqrt{2}\pi^{2}}m_{1}^{2}
\sum_{l}U_{il}U_{ml}^{*}(\frac{9}{8}\frac{m_{l}^{2}}{M_{W}^{2}}-
3\frac{m_{l}^{2}}{M_{W}^{2}}ln\frac{M_{W}}{m_{l}}), \nonumber \\
F_{1}^{R}(0)_{i \neq m}\simeq -\imath\frac{eG_{F}}{4\sqrt{2}\pi^{2}}m_{1}m_{2}
\sum_{l}U_{il}U_{ml}^{*}(-\frac{5}{8}\frac{m_{l}^{2}}{M_{W}^{2}}+
3\frac{m_{l}^{2}}{M_{W}^{2}}ln\frac{M_{W}}{m_{l}}) ,
\end{eqnarray}

and at the same time (see Eq.(10.27) in the book of Ref.[1])

\begin{eqnarray}
F_{2}^{L}(0)_{i \neq m}\simeq -\imath\frac{eG_{F}}{4\sqrt{2}\pi^{2}}m_{2}
\sum_{l}U_{il}U_{ml}^{*}(\frac{3}{4}\frac{m_{l}^{2}}{M_{W}^{2}}),
 \nonumber \\
F_{2}^{R}(0)_{i \neq m}\simeq -\imath\frac{eG_{F}}{4\sqrt{2}\pi^{2}}m_{1}
\sum_{l}U_{il}U_{ml}^{*}(\frac{3}{4}\frac{m_{l}^{2}}{M_{W}^{2}}).
\end{eqnarray}

It is straightforward to evaluate the rate:

\begin{eqnarray*}
\Gamma(f_{1}(m_{1})\rightarrow f_{2}(m_{2})+\gamma)=\frac{m_{1}^{2}-m_{2}^{2}}
{4\pi m_{1}^{2}}[2 p^{2}(p+\sqrt{p^{2}+m_{2}^{2}})(|F_{2}^{V}|^{2}
+|F_{2}^{A}|^{2})  \\
+|F_{1}^{V}|^{2}(\sqrt{p^{2}+m_{2}^{2}}-m_{2})+|F_{1}^{A}|^{2}
(\sqrt{p^{2}+m_{2}^{2}}+m_{2})-m_{2} p (F_{1}^{V}F_{2}^{V*}+ 
F_{1}^{V*}F_{2}^{V}  \\
-F_{1}^{A}F_{2}^{A*}-F_{1}^{A*}F_{2}^{A})
+p (p+\sqrt{p^{2}+m_{2}^{2}})(F_{1}^{V}F_{2}^{V*}+F_{1}^{V*}F_{2}^{V}+
F_{1}^{A}F_{2}^{A*}+F_{1}^{A*}F_{2}^{A})] ,\\
where\ p=\frac{m_{1}^{2}-m_{2}^{2}}{2 m_{1}},\ F_{i}^{V,A}=\frac
{F_{i}^{R}\pm F_{i}^{L}}{2}.  
\end{eqnarray*}

Thus, the contributions from the $F_{1}^{L,R}(\nu_{1}\neq \nu_{2})$ 
and $F_{2}^{L,R}(\nu_{1}\neq \nu_{2})$ form factors to 
the rate of $\nu_{1}\rightarrow \nu_{2}+\gamma$ (or $\mu\rightarrow e+
\gamma$)
 are of the same order of magnitude \cite{Moha}(however the neutrino
flavour diagonal
charges vanish $F^{L,R}_{1}(0)_{\nu_{i}}=0$).

To conclude, one can say that if perturbation theory is correctly applied
to mixed fermion states, one has to calculate all $F_{1,2}^{L,R}(f_{1} \neq
 f_{2})$ 
form factors to evaluate decay amplitudes in any electroweak model with lepton mixing. 
In our calculation we respect Lorentz, gauge and BRST symmetries, as well as
the renormalization conditions for mixed fermion states. A correct evaluation
 of the neutrino lifetime for certain electroweak models could be of great
 importance in the theoretical cosmology and astrophysics \cite{Moha,Kolb},
 with Sciama's decaying neutrino hypothesis as an example \cite{Sci}.


\newpage
\begin{thebibliography}{300}

\bibitem{Moha} Lee B. W. and Shrock R. E., {\sl Phys. Rev. D}, {\bf 16} (1977) 
1444; 
 Mohapatra R. N. and Pal P. B., {\sl Massive neutrinos in physics
and astrophysics} (World Scientific, Singapore) 1991, and references therein.
\bibitem{Aoki} Aoki K. et al, {\sl Suppl. of the Prog. Th. Phys.}, {\bf 73} 
 (1982) 1.
\bibitem{Boehm} B\"{o}hm M., Spiesberger H. and Hollik W., {\sl Fort. der Phys.},
{\bf 34} (1986) 687.
\bibitem{Kolb} Kolb E. W. and Turner M. S., {\sl The Early Universe} 
(Addison-Wesley Pub. Co., California, US) 1990.
\bibitem{Sci} Sciama D. W., {\sl Nature}, {\bf 346} (1990) 40;
 Sciama D. W., {\sl Modern Cosmology and the Dark Matter Problem}
 (Cambridge Univ. Press, UK) 1995.

\end{thebibliography}
\end{document}